\documentclass[letterpaper,english,reprint, aps]{revtex4-1}
\usepackage[T1]{fontenc}
\usepackage[latin9]{inputenc}
\usepackage{color}
\usepackage{babel}
\usepackage{float}
\usepackage{amsmath}
\usepackage{amssymb}
\usepackage{stmaryrd}
\usepackage{graphicx}
\usepackage{setspace}
\usepackage[unicode=true,pdfusetitle,
 bookmarks=true,bookmarksnumbered=false,bookmarksopen=false,
 breaklinks=false,pdfborder={0 0 0},pdfborderstyle={},backref=false,colorlinks=true]
 {hyperref}
\hypersetup{
 citecolor=blue}

\makeatletter

\pdfpageheight\paperheight
\pdfpagewidth\paperwidth

\usepackage{array, amsmath}
\usepackage{tabularx}

\makeatother

\begin{document}

\title{An alternative formalism for modeling spin}

\author{Sam Powers}

\affiliation{HEPCOS, Department of Physics, SUNY at Buffalo, Buffalo, NY 14260-1500,
USA}

\author{Dejan Stojkovic}

\affiliation{HEPCOS, Department of Physics, SUNY at Buffalo, Buffalo, NY 14260-1500,
USA}

\date{\today}

\begin{abstract}
We present an alternative formalism for modeling spin. The ontological
elements of this formalism are base-2 sequences of length $n$.
The machinery necessary to model physics is then developed by considering
correlations between base-2 sequences. Upon choosing a reference base-2 sequence,
a relational system of numbers can be defined, which we interpret as quantum numbers. Based on the properties of these relational quantum
numbers, the selection rules governing interacting spin systems are
derived from first principles. A tool for calculating the associated
probabilities, which are the squared Clebsch-Gordan coefficients in
quantum mechanics, is also presented. The resulting model offers a
vivid information theoretic picture of spin and interacting spin systems.
Importantly, this model is developed without making any assumptions
about the nature of space-time, which presents an interesting opportunity
to study emergent space-time models.
\end{abstract}

\maketitle

\section{Introduction}
\label{sec:introduction}

Historically, discovering new ways of obtaining established results has been an effective means of making progress in physics. The epitome of this is Hamilton's reformulation of Newtonian mechanics. Superficially, a reproduction of a known result, especially one obtained nearly a century ago, seems inconsequential. Though occasionally, as was the case for Hamilton's insight, the manner in which the old result is reproduced can open new avenues of thought and exploration. It is in this vein that we present here an alternative formalism for modeling spin, which emerges upon the consideration of two point correlations between base-2 sequences. The information theoretic roots of this alternative formalism paint a completely new picture of the conceptually elusive, but physically ubiquitous quantity known as spin. 

The information we obtain about physical systems requires measurement,
which inevitably involves one or more quantum mechanical interactions
\cite{Shannon19621949,Stone2015,Pierce1980}. While one cannot say
with certainty if interactions in nature are discrete or continuous
at the fundamental level, the observable outcome of any interaction
is always discrete. For this reason, the results of any conceivable
physical experiment can be reduced to counting. This fact stands in
stark contrast with the uncountable sets universally employed by modern theories, which are based on continuous functions satisfying
differential equations. This tension between the countable
nature of empirical data and the uncountable sets that form the foundations
of modern theories is not simply a matter of improving precision
or collecting more data \cite{Dauben1979}. Rather, it exists because
of a fundamental difference between our experience of the physical
world and the theories we use to model those experiences. This simple
observation leads us to the following quote from Niels Bohr : 
\begin{quote}
``It is wrong to think that the task of physics is to find out how
nature is. Physics concerns what we can say about nature.''\cite{Petersen1963}
\end{quote}
If we cannot prove that nature is continuous, then perhaps we should
explore theories which do not require it be so. The quantum revolution
of the twentieth century was a direct consequence of the observed
discreteness of interactions \cite{Einstein1905}. However, quantum
mechanics (QM) was built with the classical Hamiltonian in mind \cite{Neumann2018}.
This approach resulted in a strong dependence of the theory on uncountable
sets. While quantum gravity is generally considered to be the final
piece of the quantum revolution \cite{Gibbs1995,Surya_2019,RovelliCarlo1998LQG,Mukhi_2011,Loll2019},
there remain significant questions regarding the nature of the quantum
state in QM \cite{Hardy2012,Hossenfelder_2020,Ballentine1970,fuchs2010qbism,Smolin_2012}.
This less appreciated use of uncountable sets in physics was a primary
motivation for the development of the alternative formalism presented
here, which has the ability to reproduce predictions from QM under
a continuum limit, while also revealing important geometric properties
and selection rules in the finite regime. 

For nearly a century, there has been a perpetual debate regarding the reality of the quantum
state in QM. \cite{Harrigan_2010,Leifer_2014,Pusey_2012}.
That is, does the quantum state represent something truly physical,
or is it epistemic? Much of this debate occurs within the context
of the standard Dirac formalism for QM, which involves Hilbert spaces,
the Schrodinger equation, the Born rule, etc. \cite{Griffiths2005,Sakurai1995-1994}.
Applying the various no-go theorems that have resulted from this debate
to an alternative formalism is not generally useful, especially when
that alternative formalism does not assume a preexisting space-time,
as will be the case here. However, even within epistemic interpretations
of the quantum state, there is still some notion of an ontic state,
where the quantum state is simply an ensemble of these ontic states.
This conceptual picture of the quantum state is precisely the one
that develops within the formalism to be introduced here, where ontic
states are modeled by sequences of finite group elements, beginning
with the group $Z_{2}$. The information stored in the ordering of
these finite group elements is then hidden, or coarse-grained away,
leading to non-determinism in the resulting model.

While non-determinism is certainly a central feature of QM, one should
not lose sight of the profound role determinism plays in nature. As
one might imagine, incorporating the correct non-deterministic and
deterministic features into a single cohesive model for spin is no
small task. Yet, the formalism to be introduced here manages this
feat quite naturally. For example, quantities like total spin, which
is an emergent and relational property of two point correlations between
base-2 sequences, can be conserved by considering permutations of
the underlying sequences. The selection rules obeyed by interacting
spin systems can be recovered by considering three point correlations
between base-2 sequences, along with simple arithmetic arguments.
Of course, these selections rules include deterministic equations
associated with the conservation of angular momentum within interacting
spin systems. Thus, important laws of nature arise naturally within this formalism, rather than being asserted through axioms or principles. 

The probability coefficients obeyed by interacting spin systems, which
are the squared Clebsch-Gordan coefficients in QM, represent an important
test case for the development of this formalism and the subsequent
model. As previously mentioned, we make no assumptions about the nature
of space-time. Rather, our intention is to use calculations, such
as the probability coefficients for interacting spin systems, to guide
our development of space-time. The result of this calculation is a
simple closed form expression, coupled with a vivid conceptual picture
which involves two observers, one associated with each of
the constituents involved in a spin interaction experiment. These
observers, which we call Alice and Bob, each construct their own epistemic
ensemble, which encodes the knowledge each has about the physical
systems involved in the experiment. The probability coefficients are
then found by counting paths between their ensembles, such that certain
quantum numbers are conserved.

In recent decades, several serious research efforts have been made
towards producing an alternative to QM \cite{Adler2004,Hooft2014,Spekkens2004,Chiribella2010,Rovelli_1996,Hardy2001,Palmer2020,Chang2019}. Through unique combinations of motivations, development strategies,
and results, each of these efforts have contributed significantly
to a shifting paradigm, at least within
the small community of active researchers in this field. For those
familiar with these efforts, the existence of a theory beyond QM is
not some faint notion, but a plausible and attainable reality. Given
the foundational role that QM plays in science and technology, as
well as the considerable challenges facing these fields today, the
pace of scientific discourse regarding this matter must increase.
What differentiates the formalism presented here from these previous
efforts is its unique combination of simplicity and modeling power.
With a small number of mathematical tools, it has the
ability to produce the selection rules and probability coefficients
associated with a real experiment, while refraining from making any
assumptions about the nature of space-time. In other words, the formalism
and subsequent model introduced here not only offers an interesting
information theoretic picture of the quantum state as well as interactions,
but it also has clear predictive power and the potential to inform
important next steps in the development of an emergent space-time.

This paper is broken into six sections, including the introduction.
In section \ref{sec:Binary-Sequences}, the foundations of the alternative formalism will be introduced, which involves base-2 sequences and
correlations between them. In section \ref{sec:Quantum-Numbers}, our definition
of quantum numbers will be introduced, along with the notation necessary
to label sequences, or sets of sequences, using these quantum numbers.
The properties of these quantum numbers are then explored
in section \ref{sec:Selection-Rules}, which leads to the derivation
of the selection rules for interacting spin systems. The probabilities
associated with interacting spin systems are then calculated in section
\ref{sec: Probabilities}. Finally, the
implications of this work, as well as some ideas regarding future
work, are discussed in section \ref{sec:Discussion}.

\section{Sequences \label{sec:Binary-Sequences}}

The building block of this formalism is the base-2 sequence. A base-2
sequence is a list comprised of two distinct symbols, where the symbols
may be repeated and order matters. The symbols used here are
$0$ and $1$, which are the members of the finite group
$Z_{2}$ \cite{Zee2016}:

\begin{equation}
\left(\begin{array}{c}
1\\
0\\
0\\
1\\
0\\
1
\end{array}\right)\label{eq:Binary Sequence}
\end{equation}

A base-2 sequence can be of any length, which is denoted as $n$.
For a given length $n$, there will be $2^{n}$ unique sequences.
The set of all such sequences is denoted as $S^{1}(n)$. In physics, it is well known that most of the information
contained in a composite system does not lie in its subsystems, but
actually in the correlations between its subsystems \cite{Page1993}.
For this reason, we are motivated to introduce the set $S^{2}(n)$,
which is the set of all two point correlations between base-2 sequences.
An element of the set $S^{2}(4)$ is given here:

\begin{equation}
\left(\begin{array}{cc}
1 & 0\\
0 & 1\\
1 & 1\\
1 & 0
\end{array}\right)\in S^{2}(n=4)\label{eq: element of S^2}
\end{equation}

Individual elements of these sets are denoted as $s^{2}\in S^{2}(n)$,
where $n$ has been suppressed. Using this notation,
an element of $S^{2}(n)$ can be constructed using two elements of
$S^{1}(n)$ like so, where the $\otimes$ symbol is used to denote
the correlation operator:

\begin{equation}
s^{1}\otimes s'^{1}=s^{2}\label{eq: first product}
\end{equation}

A more explicit representation of the operation shown in equation
(\ref{eq: first product}) is given here, where a particular example
of $s^{1}$ and $s'^{1}$ has been chosen:

\begin{equation}
\left(\begin{array}{c}
1\\
0\\
0\\
1\\
0\\
1
\end{array}\right)\otimes\left(\begin{array}{c}
0\\
0\\
1\\
1\\
0\\
0
\end{array}\right)=\left(\begin{array}{cc}
1 & 0\\
0 & 0\\
0 & 1\\
1 & 1\\
0 & 0\\
1 & 0
\end{array}\right)\label{eq:Binary Matrix}
\end{equation}

Simply put, base-2 sequences are the bricks of this formalism, while
the correlation operator is the mortar. This correlation operation
can also be thought of as an increase in basis. While an element of
$S^{1}(n)$ is a sequence written in base-2, elements of $S^{2}(n)$
can be thought of as sequences written in base-4, where the new basis
 elements, or symbols, are the members of the group $Z_{2}\otimes Z_{2}$.
While one can always use the base-2 representation, it will be conceptually
beneficial to introduce alternative symbols for the basis elements
of $S^{2}(n)$; $00=A$, $11=B$, $10=C$, and $01=D$. With this
notation in hand, equation (\ref{eq:Binary Matrix}) can be rewritten
as follows:

\begin{equation}
\left(\begin{array}{c}
1\\
0\\
0\\
1\\
0\\
1
\end{array}\right)\otimes\left(\begin{array}{c}
0\\
0\\
1\\
1\\
0\\
0
\end{array}\right)=\left(\begin{array}{c}
C\\
A\\
D\\
B\\
A\\
C
\end{array}\right)\label{eq:Binary Matrix base 4}
\end{equation}

\begin{figure}

\resizebox{0.45\textwidth}{!}{
  \includegraphics{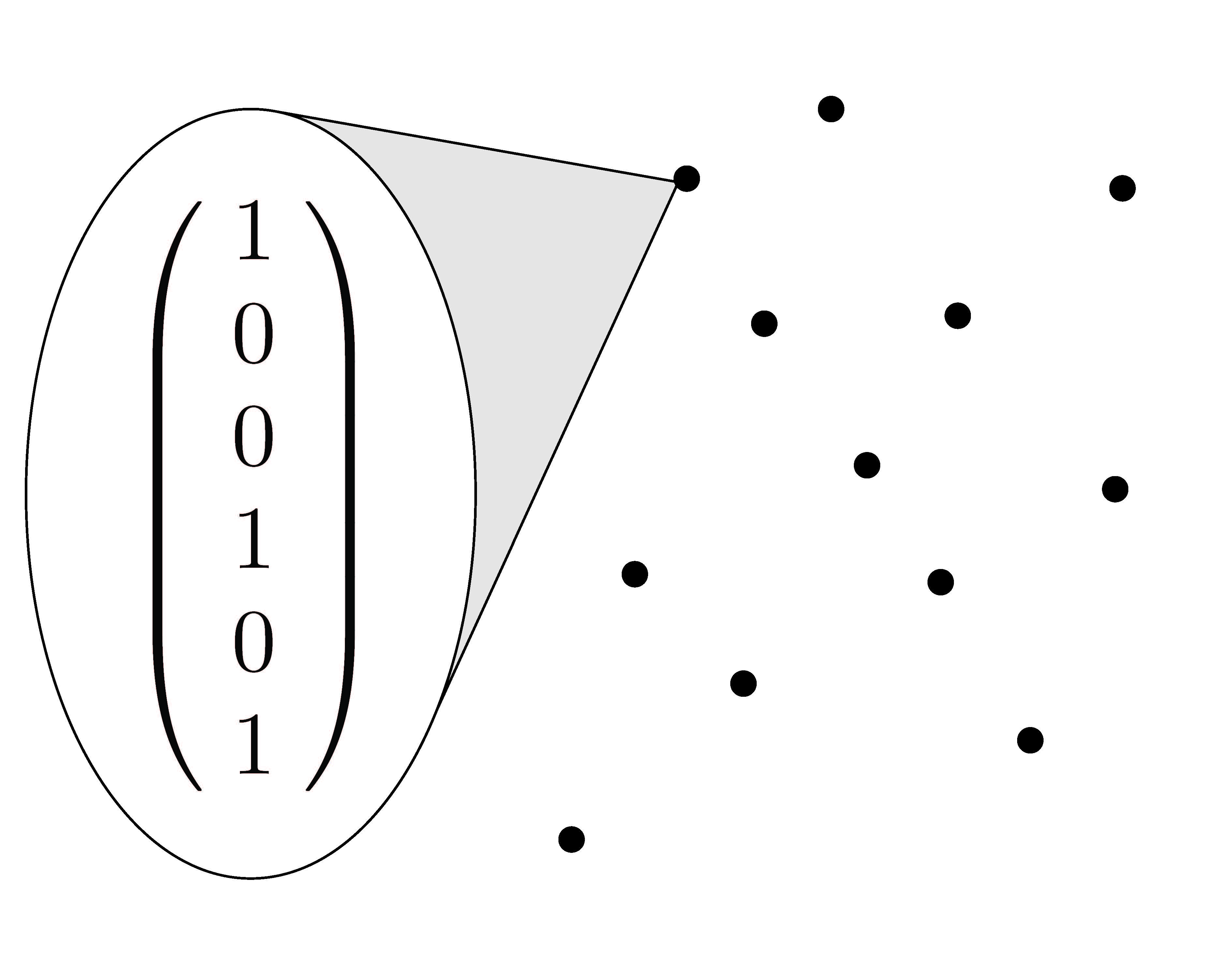}
}

\caption{A random selection of base-2 sequences are visualized as points in
some abstract space, where the position of each point has no physical
significance.}
\label{fig: base-2 example}       
\end{figure}

More generally, the approach taken in this formalism is to construct random
base-2 $n\times d$ matrices by gluing together $d$ base-2 sequences
of length $n$ using the $\otimes$ operator. It may be useful to
imagine each base-2 sequence as a point in some abstract space. Though
the details of that space, as well as the distribution of the points
within it, have no physical significance just yet (Figure \ref{fig: base-2 example}).
The distribution of points within this abstract space is related to
the issue of ordering sets. Given a set of base-2 sequences,
which one should come first? Binary languages in computer science
offer perfectly reasonable answers to this question. However, those
approaches to ordering base-2 sequences rely on the information stored
in the ordering of the base-2 basis elements, which we plan to hide,
or coarse-grain away. In the following section, an ordering scheme
will be introduced that can survive such a step.

\section{Quantum Numbers\label{sec:Quantum-Numbers}}

Within this model, the information stored in the configuration of
the basis elements comprising a sequence is hidden, or coarse-grained
away. This means that the successful ordering scheme will only be
a partial ordering of base-2 sequences, rather than a total ordering.
This is an essential feature of this formalism, which leads directly to
non-determinism in the subsequent model. 

The ordering scheme employed here requires the introduction of a relational
system of numbers. This number system is a function of the reference
sequence, which is a particular base-2 sequence chosen from the set
$S^{1}(n)$, and is denoted as $s_{0}^{1}$. Using the correlation
operation, this reference sequence is then used to construct elements
of the set $S^{2}(n)$, which are base-4 sequences. The number of
times a particular basis element appears in a sequence is called a
count. For each relationship between the reference sequence and another
element of $S^{1}(n)$, there are four associated counts. These four
counts are denoted as $\tilde{A}$, $\tilde{B}$, $\tilde{C}$, and
$\tilde{D}$, where the tilde notation has been introduced to distinguish
each count from its associated base-4 basis element.

From these four counts, a relational set of measures can be defined
for each base-4 sequence, which we interpret as quantum numbers. It will be shown that the quantum numbers $j=\frac{\tilde{C}+\tilde{D}}{2}$
and $m=\frac{\tilde{C}-\tilde{D}}{2}$ share important properties
with total spin and the z-component of spin,
respectively \cite{Griffiths2005}. Moreover, the quantum number
$j$, which is closely related to the Hamming distance in computer
science, is a metric. This means that for any choice of three base-2
sequences, one can be placed at each of the vertices of a triangle,
where $j$ is the length of the edge connecting two vertices. This
feature endows this formalism with important geometric properties.

A complete set of quantum numbers allows one to determine the number
of times each basis element appears within a particular sequence.
To make $j$ and $m$ complete, the quantum numbers $g=\frac{\tilde{A}+\tilde{B}}{2}$
and $l=\frac{\tilde{A}-\tilde{B}}{2}$, which do not yet have established
physical analogues, must be included. Thus, the complete set of quantum
numbers for a particular base-4 sequence is as follows:

\noindent\begin{tabularx}{\columnwidth}{@{}XX@{}}
\begin{equation}
j=\frac{\tilde{C}+\tilde{D}}{2}\label{eq: j}
\end{equation}
&
\begin{equation}
m=\frac{\tilde{C}-\tilde{D}}{2}\label{eq:m}
\end{equation}
\end{tabularx}

\noindent\begin{tabularx}{\columnwidth}{@{}XX@{}}
\begin{equation}
g=\frac{\tilde{A}+\tilde{B}}{2}\label{eq: g}
\end{equation}
&
\begin{equation}
l=\frac{\tilde{A}-\tilde{B}}{2}\label{eq: l}
\end{equation}
\end{tabularx}

\noindent\begin{tabularx}{\columnwidth}{@{}XX@{}}
\begin{equation}
-j\leq m\leq j\label{eq: m range}
\end{equation}
&
\begin{equation}
-g\leq l\leq g\label{eq: l range}
\end{equation}
\end{tabularx}

The quantum numbers defined in equations (\ref{eq: j}-\ref{eq: l})
will serve as ordering parameters. Notationally, these ordering parameters
can be used to distinguish one set of sequences from another. In the
case of base-2 sequences, the subset of $S^{1}(n)$ containing all
base-2 sequences with the quantum numbers $j$, $m$, $g$, and $l$,
as determined by the chosen reference sequence $s_{0}^{1}$, is denoted
as follows: $S^{1}(j,m,g,l)\subset S^{1}(n)$. Note that $n=2j+2g=\tilde{A}+\tilde{B}+\tilde{C}+\tilde{D}$,
making explicit mention of $n$ unnecessary if both $j$ and $g$
are given. An element of the subset $S^{1}(j,m,g,l)$ can then be
denoted by including subscripts like so: $s_{j,m,g,l}^{1}\in S^{1}(j,m,g,l)$.
With this notation in hand, the correlation operation can be defined
as follows: 

\begin{equation}
s_{0}^{1}\otimes s_{j,m,g,l}^{1}=s_{j,m,g,l}^{2}\label{eq: correlation}
\end{equation}

Equation (\ref{eq: correlation}) raises an important issue, which
is that the quantum numbers $j$, $m$, $g$, and $l$ can be used
to label base-2 sequences like $s_{j,m,g,l}^{1}$, as well as base-4
sequences like $s_{j,m,g,l}^{2}$. When used to label base-2 sequences,
these quantum numbers are functions of the chosen reference sequence
$s_{0}^{1}$, resulting in a relational ordering scheme. This just
means that the quantum numbers $j$, $m$, $g$, and $l$ associated
with a particular base-2 sequence may vary depending on the reference
sequence. On the other hand, the subset of base-4 sequences associated
with $j$, $m$, $g$, and $l$ will include all possible two point
correlations between base-2 sequences that result in those quantum
numbers. 

\begin{figure}

\resizebox{0.45\textwidth}{!}{
  \includegraphics{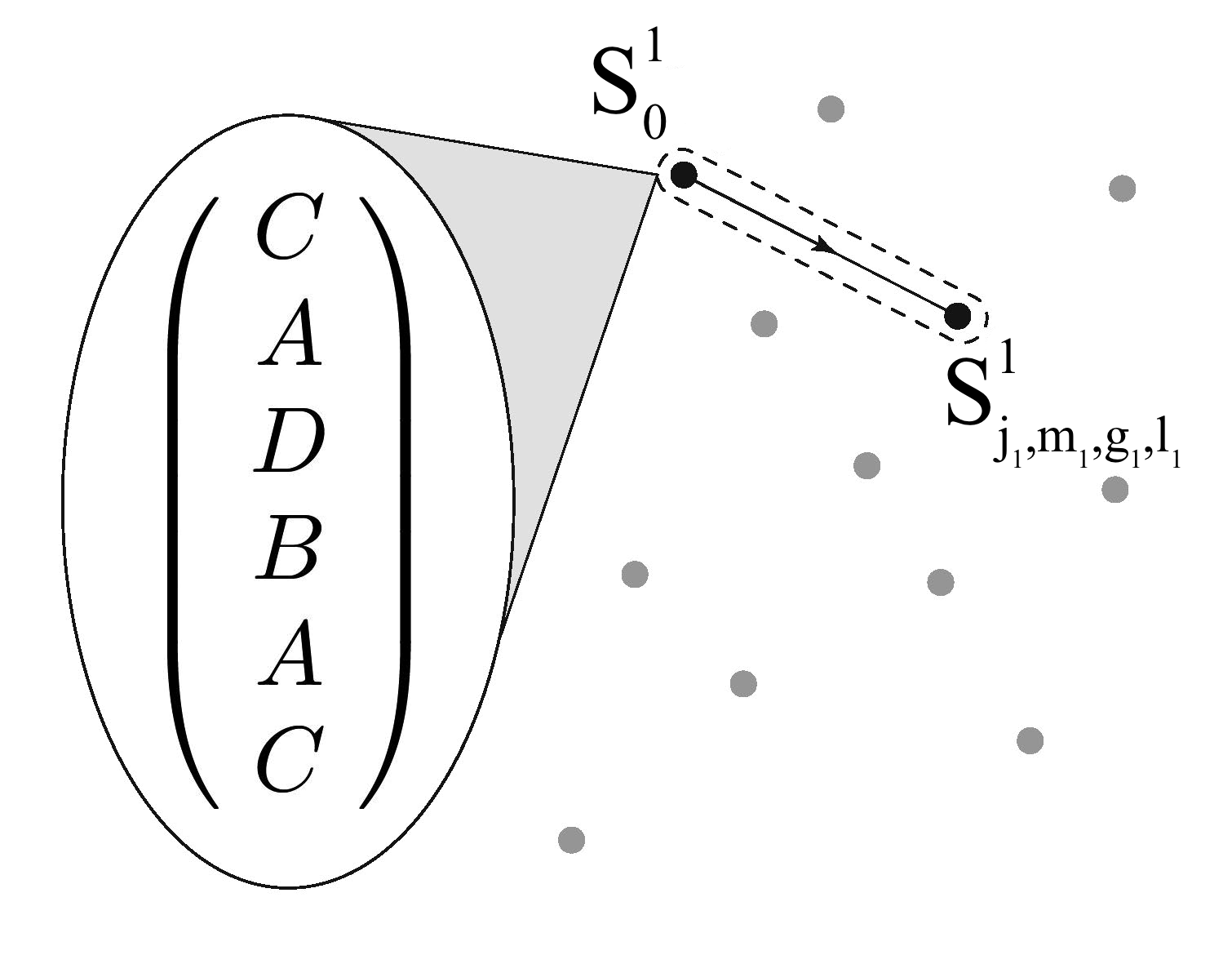}
}

\caption{A correlation of two base-2 sequences, which is a base-4 sequence
and an element of $S^{2}(n)$, can be visualized as a directed edge
connecting two vertices. By choosing a reference sequence, we can
assign quantum numbers to the remaining base-2 sequences and order
the points introduced in Figure \ref{fig: base-2 example} accordingly.
For the case shown here, the quantum numbers are $j_{01}=\frac{3}{2}$,
$m_{01}=+\frac{1}{2}$, $g_{01}=\frac{3}{2}$, and $l_{01}=+\frac{1}{2}$.}
\label{fig: Network}       
\end{figure}

On a more technical note, the position of the reference sequence within
the correlation shown in equation (\ref{eq: correlation}) is important
due to the asymmetry of the $C=10$ and $D=01$ basis elements under
the commutation operation. Under this operation, the counts $\tilde{C}$
and $\tilde{D}$ are exchanged, implying the quantum number $m$ must
change sign according to equation (\ref{eq:m}). Notationally, subscripts
can be added to each quantum number to convey the orientation of the
correlation like so: $s_{0}^{1}\otimes s_{j_{1},m_{1},g_{1},l_{1}}^{1}=s_{j_{01},m_{01},g_{01},l_{01}}^{2}$.
Again, the only quantum number that changes sign under the exchange
of these indices is $m$: $m_{01}=-m_{10}$. The picture associated
with the operation in equation (\ref{eq: correlation}), which can
be visualized as a directed edge connecting two vertices, is given
in Figure \ref{fig: Network}. 

The physical interpretation of the operation shown in equation (\ref{eq: correlation})
is a single measurement. We read the expression $s_{0}^{1}\otimes s_{j_{1},m_{1},g_{1},l_{1}}^{1}=s_{j_{01},m_{01},g_{01},l_{01}}^{2}$
as follows: the sequence to the left of the $\otimes$ symbol ``looks''
at the sequence to the right and ``sees'' the quantum numbers $j$,
$m$, $g$, and $l$. Importantly, what the reference sequence ``sees''
is not actually the other base-2 sequence, but rather the coarse-grained
relationship between the sequences. From this picture, an interesting
question arises. Given two base-2 sequences with the quantum numbers
$j$, $m$, $g$, $l$ and $j'$, $m'$, $g'$, $l'$, as determined
by a common reference sequence, which quantum numbers describe their
relationship? As will be shown in the following section, the answer
to this question contains the selection rules for interacting spin
systems. 

\section{Selection Rules\label{sec:Selection-Rules}}

In this section, a single reference sequence is used to determine
the quantum numbers $j$, $m$, $g$, and $l$ for two different base-2
sequences. Independently, these operations take the following form,
where the choice of indices will be discussed shortly: 
\begin{equation}
s_{j_{1},m_{1},g_{1},l_{1}}^{1}\otimes s_{0}^{1}=\begin{array}{c}
s_{j_{10},m_{10},g_{10},l_{10}}^{2}\end{array}\label{eq: measurement 1}
\end{equation}
\begin{equation}
s_{0}^{1}\otimes s_{j_{2},m_{2},g_{2},l_{2}}^{1}=\begin{array}{c}
s_{j_{02},m_{02},g_{02},l_{02}}^{2}\end{array}\label{eq: measurement 2}
\end{equation}

Using the quantum numbers $j_{10}$, $m_{10}$, $g_{10}$, $l_{10}$,
$j_{02}$, $m_{02}$, $g_{02}$, and $l_{02}$, we can infer some properties of the following relationship:

\begin{equation}
s_{j_{1},m_{1},g_{1},l_{1}}^{1}\otimes s_{j_{2},m_{2},g_{2},l_{2}}^{1}=\begin{array}{c}
s_{j_{12},m_{12},g_{12},l_{12}}^{2}\end{array}\label{eq: measurement 3}
\end{equation}

Recall that the ordering of the indices on each quantum number only
impacts the sign of $m$. The choice of index orderings
in equations (\ref{eq: measurement 1}-\ref{eq: measurement 3}) has
been made for pedagogical reasons, but any other ordering is equally
valid (there are eight unique choices). By simple arguments (see appendix
\ref{sec:Derivation-of-AMRA}), we can prove the following relationships
between the quantum numbers $j_{10}$, $m_{10}$, $g_{10}$, $l_{10}$,
$j_{02}$, $m_{02}$, $g_{02}$, and $l_{02}$ and $j_{12}$, $m_{12}$,
$g_{12}$, and $l_{12}$, where it is assumed that $n\geq2j_{10}+2j_{02}$:

\begin{equation}
n=2(j_{10}+g_{10})=2(j_{02}+g_{02})\label{eq: n}
\end{equation}
\begin{equation}
m_{12}=m_{10}+m_{02}=l_{02}-l_{10}\label{eq: M-1}
\end{equation}
\begin{equation}
l_{12}=l_{10}+m_{02}=l_{02}-m_{10}\label{eq: L-1}
\end{equation}
\begin{equation}
|j_{10}-j_{02}|\leq j_{12}\leq j_{10}+j_{02}\label{eq: J range}
\end{equation}
\begin{equation}
\frac{n}{2}-j_{10}-j_{02}\leq g_{12}\leq\frac{n}{2}-|j_{10}-j_{02}|\label{eq: G range}
\end{equation}

\begin{figure}

\resizebox{0.45\textwidth}{!}{
  \includegraphics{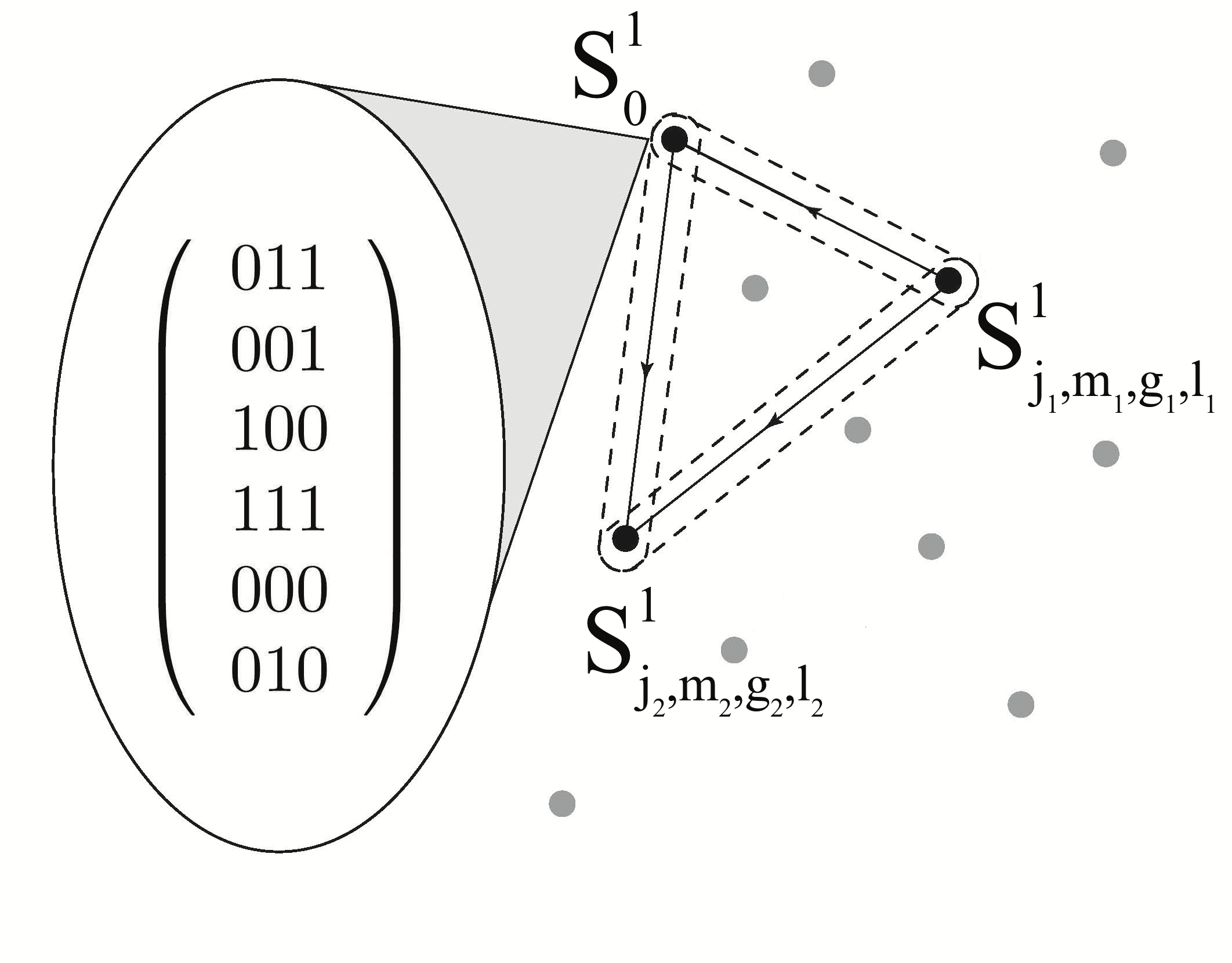}
}

\caption{A correlation of three base-two sequences, which is a base-8 sequence
and an element of $S^{3}(n)$, can be visualized as a directed graph
with three vertices and three directed edges. Among the rules governing
the quantum numbers associated with this graph are those for interacting
spin systems in QM. For the case shown here, the quantum numbers are
$j_{10}=\frac{3}{2}$, $m_{10}=-\frac{1}{2}$, $g_{10}=\frac{3}{2}$,
$l_{10}=+\frac{1}{2}$, $j_{02}=1$, $m_{02}=0$, $g_{02}=2$, $l_{02}=0$,
and $j_{12}=\frac{3}{2}$, $m_{12}=-\frac{1}{2}$, $g_{12}=\frac{3}{2}$,
$l_{12}=+\frac{1}{2}$.}
\label{fig:base-8 }      
\end{figure}

Equations (\ref{eq: m range}), (\ref{eq: M-1}), and (\ref{eq: J range})
contain the selection rules governing interacting spin systems in
QM \cite{EDMONDS1985}. Because there are three base-2 sequences
involved, the true object of interest in this section is a three point
correlation between base-2 sequences, where the set of all such correlations
is denoted as $S^{3}(n)$. As with $S^{2}(n)$, which can be interpreted
as the set of all base-4 sequences, $S^{3}(n)$ can be interpreted
as the set of all base-8 sequences, where the basis elements are members
of the group $Z_{2}\otimes Z_{2}\otimes Z_{2}$. Rather than introducing
new symbols for each of these eight basis elements, as done for base-4
sequences, the base-2 representation will be used: $000$, $111$,
$101$, $010$, $100$, $011$, $001$, and $110$. A visualization
of a three point correlation among base-2 sequences is offered in
Figure \ref{fig:base-8 }, which takes the form of a directed graph.
Based on the choice of index orderings made in equations (\ref{eq: measurement 1}-\ref{eq: measurement 3}),
the base-4 basis element associated with each of the two point correlations
of interest can be identified as follows, where $X\in\{0,1\}$: 

\[
s_{j_{10},m_{10},g_{10},l_{10}}^{2}\rightarrow\underline{XX}X
\]

\[
s_{j_{02},m_{02},g_{02},l_{02}}^{2}\rightarrow X\underline{XX}
\]

\[
s_{j_{12},m_{12},g_{12},l_{12}}^{2}\rightarrow\underline{X}X\underline{X}
\]

Because of these relations, base-8 counts can be associated with base-4
counts like so: $\tilde{C}_{10}=\widetilde{\underline{10}0}+\widetilde{\underline{10}1}$,
$\tilde{D}_{12}=\widetilde{\underline{0}0\underline{1}}+\widetilde{\underline{0}1\underline{1}}$,
etc.. This enables us to define a complete set of base-8 quantum
numbers that include base-4 quantum numbers like $j_{10}$, $m_{10}$,
$j_{02}$, and $m_{02}$. In fact, defining a complete set of base-8
quantum numbers only requires the introduction of one new quantum
number, which can also be interpreted as a count: 

\begin{equation}
k=\widetilde{010}\label{eq: k0-1}
\end{equation}

The complete set of base-8 quantum numbers to be used to label base-8
sequences are $n$, $j_{10}$, $j_{02}$, $m_{10}$, $m_{02}$, $j_{12}$,
$l_{12}$, and $k$, which are defined in Table \ref{tab:Base-8 measures},
along with $m_{12}$ for completeness. Table \ref{tab:Base-8-counts}
provides the map from quantum numbers back to base-8 counts. The definitions offered in these tables will vary depending on how
one orders the indices in equations (\ref{eq: measurement 1}-\ref{eq: measurement 3}),
though the results obtained herein hold for any choice.

\begin{table}
\begin{raggedleft}
\[
\begin{array}{ll}
k=\widetilde{010} & l_{12}=\frac{\widetilde{000}+\widetilde{010}-\widetilde{111}-\widetilde{101}}{2}\\
j_{10}=\frac{\widetilde{100}+\widetilde{101}+\widetilde{011}+\widetilde{010}}{2} & j_{02}=\frac{\widetilde{110}+\widetilde{101}+\widetilde{001}+\widetilde{010}}{2}\\
m_{10}=\frac{\widetilde{100}+\widetilde{101}-\widetilde{011}-\widetilde{010}}{2} & m_{02}=\frac{\widetilde{110}+\widetilde{010}-\widetilde{001}-\widetilde{101}}{2}\\
j_{12}=\frac{\widetilde{100}+\widetilde{110}+\widetilde{011}+\widetilde{001}}{2} & m_{12}=\frac{\widetilde{100}+\widetilde{110}-\widetilde{011}-\widetilde{001}}{2}
\end{array}
\]
$\begin{array}{l}
n=\widetilde{100}+\widetilde{011}+\widetilde{110}+\widetilde{001}+\widetilde{101}+\widetilde{010}+\widetilde{000}+\widetilde{111}\end{array}$
\par\end{raggedleft}
\caption{Base-8 Quantum Numbers}
\label{tab:Base-8 measures}
\end{table}

\begin{table}
\begin{raggedleft}
$$\begin{array}{ll}
\widetilde{010}=k & \widetilde{101}=j_{10}+j_{02}-j_{12}-k\\
\widetilde{100}=m_{10}-j_{02}+j_{12}+k & \widetilde{011}=j_{10}-m_{10}-k\\
\widetilde{110}=j_{02}+m_{02}-k & \widetilde{001}=j_{12}+k-m_{02}-j_{10}\\
\widetilde{111}=\frac{n}{2}-l_{12}-j_{10}-j_{02}+k & \widetilde{000}=\frac{n}{2}-j_{12}+l_{12}-k
\end{array}$$\\
\par\end{raggedleft}
\caption{Base-8 Counts}
\label{tab:Base-8-counts}
\end{table}

In all cases but one, the quantum numbers defined in Table \ref{tab:Base-8 measures}
can be found by collecting the three complete sets of base-4 quantum
numbers $(j_{10},m_{10},g_{10},l_{10})$, $(j_{02},m_{02},g_{02},l_{02})$
and $(j_{12},m_{12},g_{12},l_{12})$, with the only exception being
$k$. As discussed in section \ref{sec:Quantum-Numbers}, base-4 quantum
numbers arise from the operation depicted in equation (\ref{eq: correlation}),
which is interpreted as a single measurement. The fact that the base-8
quantum number $k$ cannot be determined by collecting a group of
individual measurements suggests that it is non-local within this
model, while the other seven quantum numbers $n$, $j_{10}$, $j_{02}$,
$m_{10}$, $m_{02}$, $j_{12}$, and $l_{12}$ are local. As will
be seen in the following section, the non-local quantum number $k$
will play an important role in the phenomenon of interference.

\section{Probabilities\label{sec: Probabilities}}
The physical scenario of interest in this section is one in which
a system with spin quantum numbers $(j_{12},m_{12})$ is comprised
of, or decays into two systems with spin quantum numbers $(j_{10},m_{10})$
and $(j_{02},m_{02})$. The question of interest is this: Given the
priors $j_{10}$, $j_{02}$, $j_{12}$, and $m_{12}$, what is the
probability of observing a particular combination of $m_{10}$ and
$m_{02}$? 

\begin{figure*}

\resizebox{1\textwidth}{!}{
  \includegraphics{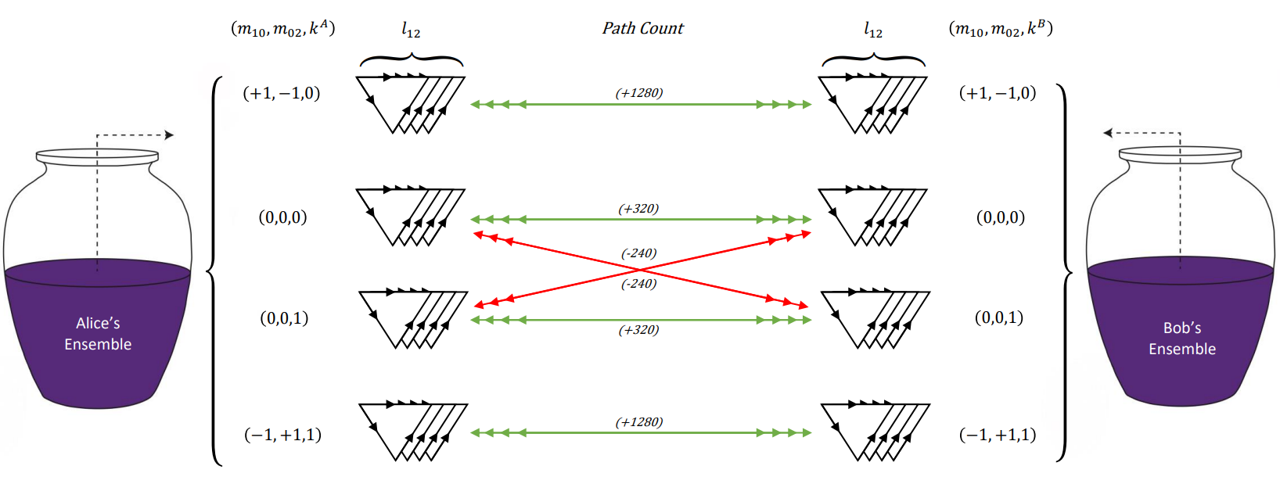}
}

\caption{An example of the path counting procedure between Alice's and Bob's
ensemble for the priors $n=6$, $j_{10}=1$, $j_{02}=1$, $j_{12}=1$,
and $m_{12}=0$. Each triangle represents a subset of $S^{3}(n)$
(the set of all base-8 sequences) with a unique combination of eight
quantum numbers. The probability of obtaining a particular combination
of $m_{10}$ and $m_{02}$ can be found by dividing the number of
paths associated with that combination by the total sum of paths for
all combinations.}
\label{fig:An-example-of}     
\end{figure*}

Answering this question within the model developed here will require
the construction of two sets of base-8 sequences, one associated with
the experiment used to collect the quantum number $m_{10}$ and one
for $m_{02}$. These two sets can be interpreted as epistemic ensembles
representing the knowledge of observers named Alice and
Bob, where Alice is responsible for collecting $m_{10}$ and Bob $m_{02}$.
The probabilities of interest can then be calculated by counting the
number of unique ways to pair base-8 sequences from Alice's ensemble
with those in Bob's, while accounting for a form of interference.
In particular, we will be interested in those pairs which share the
same combination of local quantum numbers $n$, $j_{10}$, $j_{02}$,
$m_{10}$, $m_{02}$, $j_{12}$, and $l_{12}$, where interference
is driven by the difference between Alice's and Bob's value of the
non-local quantum number $k$. 

A single pair of sequences from separate ensembles is interpreted
as a path within this model. That is, the probabilities being calculated
in this section are related to counting local quantum number conserving
paths between Alice's and Bob's ensembles. Path interference is then
driven by a measure of disagreement between Alice and Bob regarding
the value of the non-local quantum number $k$. Specifically, paths
for which $k^{B}-k^{A}$ is odd interfere destructively with those
in which it is even, where the superscript indicates which ensemble
each $k$ is associated with. For each of these paths, there is an
associated map which connects Alice's and Bob's sequences under the
addition modulo two operation (see appendix \ref{sec:Maps}). The
maps of interest in this calculation, which conserve local quantum
numbers, generate permutations of the underlying base-2 sequences. 

Now that the general framework of this calculation has been established,
all that remains is to construct Alice's and Bob's ensembles, which
will require the introduction of two combinatorial tools \cite{Faticoni2014}.
One which will simply count the number of base-8 sequences associated
with a particular combination of quantum numbers, and one that will
account for the fact that Alice and Bob are actually performing a
measurement on part of the total system. 

The number of sequences associated with a particular combination of
quantum numbers can be found by counting permutations. For base-8
sequences, this can be accomplished by using the following combinatorial
tool, where Table \ref{tab:Base-8-counts} can be used to convert
from quantum numbers to counts:

\begin{multline}
\Phi(n,j_{10},j_{02},m_{10},m_{02},j_{12},l_{12},k)=\\
\frac{n!}{\widetilde{010}!\widetilde{101}!\widetilde{100}!\widetilde{011}!\widetilde{001}!\widetilde{011}!\widetilde{000}!\widetilde{111}!}\label{eq: Base-8 counter}
\end{multline}

In cases where not all quantum numbers are known, equation (\ref{eq: Base-8 counter})
can be summed over for all possible combinations of the unknown quantum
numbers. For the calculation of interest in this section, the priors
$j_{10}$, $j_{02}$, and $j_{12}$, along with a particular combination
of $m_{10}$ and $m_{02}$ constitute five of the eight quantum numbers
necessary to qualify as complete. In addition to these, we will also
require that all sequences share a common length $n$, where the only
restriction will be that $n\geq2j_{10}+2j_{02}$.
The two remaining quantum numbers $l_{12}$ and $k$ must then be
summed over, where the bounds of these sums can be found in appendix
\ref{sec:Derivation-of-summation}. 

To account for Alice's and Bob's measurement of $m_{10}$ and $m_{02}$,
respectively, one additional combinatorial tool must be introduced. The
purpose of this tool is to modify the information encoded into the
base-8 sequences being counted by equation (\ref{eq: Base-8 counter}).
This modification pertains to the base-4 basis elements associated
with the quantum numbers $m_{10}$ and $m_{02}$, which are $(C_{10},D_{10})$
and $(C_{02},D_{02})$, respectively. This combinatorial tool takes
the following form, which has the effect of holding these base-4 basis
elements fixed when counting base-8 permutations:

\begin{equation}
F^{A}(n,j_{10},m_{10})=\frac{\tilde{C}_{10}!\tilde{D}_{10}!(n-\tilde{C}_{10}-\tilde{D}_{10})!}{n!}\label{eq: F10 base-4}
\end{equation}
\begin{equation}
F^{B}(n,j_{02},m_{02})=\frac{\tilde{C}_{02}!\tilde{D}_{02}!(n-\tilde{C}_{02}-\tilde{D}_{02})!}{n!}\label{eq: F02 base-4}
\end{equation}

For clarity, these expressions can also be written in terms of base-8
counts like so:

\begin{multline}
F^{A}(n,j_{10},m_{10})=\\
\frac{(\widetilde{101}+\widetilde{100})!(\widetilde{010}+\widetilde{011})!(\widetilde{000}+\widetilde{111}+\widetilde{110}+\widetilde{001})!}{n!}\label{eq: F10 base-8}
\end{multline}
\begin{multline}
F^{B}(n,j_{02},m_{02})=\\
\frac{(\widetilde{010}+\widetilde{110})!(\widetilde{101}+\widetilde{001})!(\widetilde{000}+\widetilde{111}+\widetilde{100}+\widetilde{011})!}{n!}\label{eq: F02 base-8}
\end{multline}

For a particular combination of $m_{10}$ and $m_{02}$,
the number of local quantum number conserving paths between Alice's
and Bob's ensembles, while accounting for interference, is given by
the following expression, where we have suppressed all arguments not being summed over:

\begin{multline}
\Upsilon(n,j_{10},j_{02},m_{10},m_{02},j_{12})=\\
\sum_{k^{A},k^{B}}\sum_{l_{12}}(-1)^{(k^{B}-k^{A})}\Phi(l_{12},k^{B})F^{B}\Phi(l_{12},k^{A})F^{A}\label{eq: upsilon}
\end{multline}

The closed form expression for calculating the probability of observing
a particular combination of $m_{10}$ and $m_{02}$ is as follows,
where the normalization is simply equation (\ref{eq: upsilon}) summed
over the allowed combinations of $m_{10}$ and $m_{02}$, given the
prior $m_{12}$:

\begin{multline}
P(m_{10},m_{02}|n,j_{10},j_{02},j_{12},m_{12})=\\
\frac{\Upsilon(n,j_{10},j_{02},m_{10},m_{02},j_{12})}{\sum_{m_{1},m_{2}}\Upsilon(n,j_{10},j_{02},m_{10},m_{02},j_{12})}\label{eq: probabilities}
\end{multline}

A depiction of the calculation associated with equation (\ref{eq: probabilities})
is offered in figure \ref{fig:An-example-of}, in which a sample calculation
is performed. The priors associated with this sample calculation are
$n=6$, $j_{10}=1$, $j_{02}=1$, $j_{12}=1$, and $m_{12}=0$. Given
these priors, along with equations (\ref{eq: m range}) and (\ref{eq: M-1}), the three allowed combinations of $m_{10}$ and $m_{02}$ are $(+1,-1)$, $(0,0)$, and $(-1,+1)$. By summing over the paths depicted in figure \ref{fig:An-example-of}, the probability of obtaining a particular combination of $m_{10}$ and $m_{02}$ is as follows:

\[
P(+1,-1|6,1,1,1,0)=\frac{1280}{2720}=0.470588
\]
\[
P(0,0|6,1,1,1,0)=\frac{160}{2720}=0.058824
\]
\[
P(-1,+1|6,1,1,1,0)=\frac{1280}{2720}=0.470588
\]

\begin{figure}

\resizebox{0.45\textwidth}{!}{
  \includegraphics{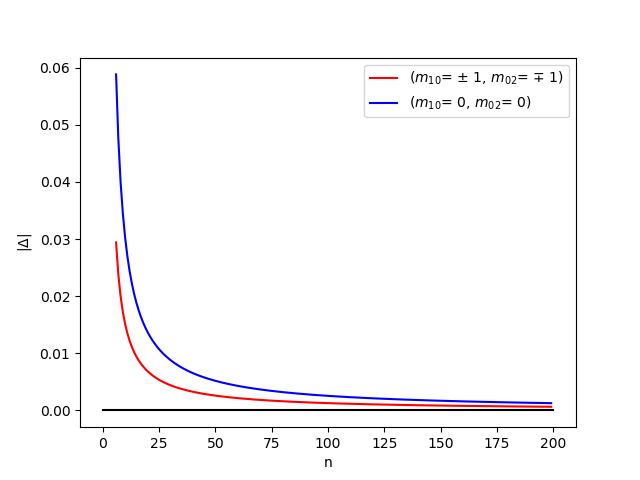}
}

\caption{The magnitude of the difference between the prediction yielded by
equation (\ref{eq: probabilities}) and that of QM, plotted as a function
of the sequence length $n$, where $|\Delta|=|P\left(m_{10},m_{02}|n,j_{10},j_{02},j_{12},m_{12}\right)-\left\langle j_{1},j_{2},m_{1},m_{2}|j_{1},j_{2},J,M\right\rangle ^{2}|$.}
\label{fig:The-magnitude-of}     
\end{figure}

The difference between these predictions and those of QM, which are
$0.5$, $0.0$, and $0.5$ for $(+1,-1)$, $(0,0)$, and $(-1,+1)$,
respectively, are plotted as a function of $n$ in figure \ref{fig:The-magnitude-of}.
The deviation between the predictions of this model and that of QM
can be made arbitrarily small by increasing $n$. In the limit that
$n$ goes to infinity, the number of sequences in Alice's and Bob's
ensembles becomes uncountable. While this model cannot be falsified
by studying deviations from QM, proving that $n$ is finite is certainly
possible.

Within Dirac's formalism for QM, the primary method of calculating
these probabilities, which are the squared Clebsch-Gordan coefficients,
is a recursive algorithm employing ladder operators. There is also
a more technical derivation associated with tensor decomposition,
which requires a background in representation theory. Regardless of
the method of derivation, there is a closed form, or non-recursive
method of calculating the square roots of these probabilities. This
expression, which is equation (\ref{eq: Original CGs}) in appendix\textbf{
}\ref{sec:Expansion-of-the}, is equivalent to equation (\ref{eq: probabilities})
in the limit of large $n$ (figure \ref{fig:The-magnitude-of}). Beyond
issues of aesthetics, equation (\ref{eq: Original CGs}) also lacks
any clear explanatory power within QM. For example, its not even obvious
that it is a probability, whereas equation (\ref{eq: probabilities})
clearly takes the form of a frequency. Finally, the method of calculating
probabilities by counting paths between two epistemic ensembles appears
to be a far more general framework than this particular calculation.
One is free to encode a wide variety of physical scenarios into this
scheme, which is of significant interest for future work.

\section{Discussion \label{sec:Discussion}}

Why should spin be the focus of an alternative formalism for modeling quantum mechanical systems? We can certainly make a case that spin is among the most fundamental
features of physical systems. Spin is even used as a building
block for space-time itself \cite{penrose1971angular,Rovelli1995}.
However, the truth is that a model for spin
was not the original objective of this research effort. Instead, it
began as a deductive approach to discretizing the quantum state in
QM, in which the starting point was the set of all base-2 sequences
of length $n$. By considering two and three point correlations between
the elements of the set $S^{1}(n)$, or the set of all base-2 sequences
of length $n$, a relational set of quantum numbers emerged. The selection rules and probabilities for interacting spin systems then developed naturally by asking simple questions of the resulting formalism. Though a model for spin was not the original objective of this research effort, the manner in which it emerged is striking. 

\begin{figure}

\resizebox{0.45\textwidth}{!}{
  \includegraphics{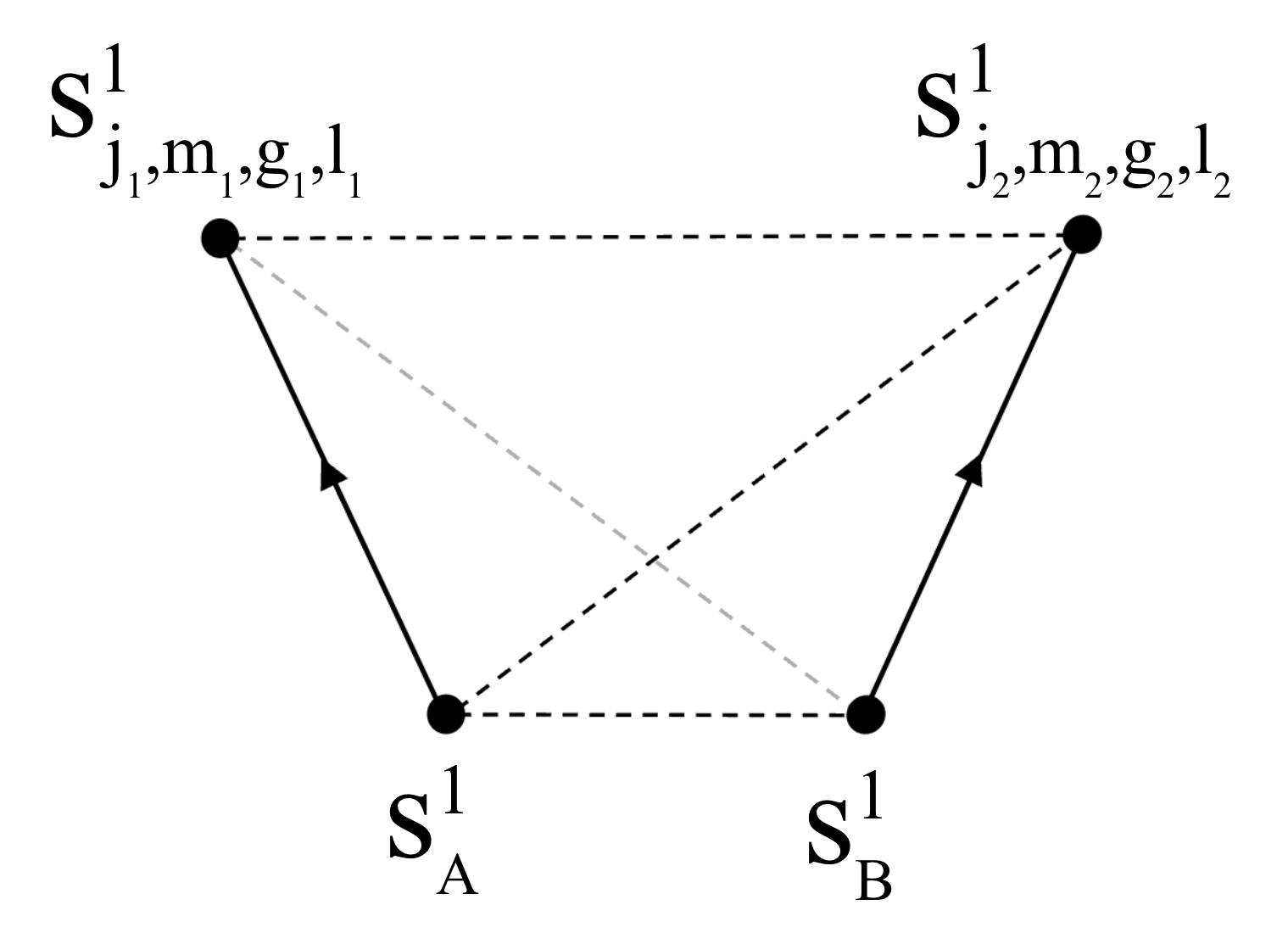}
}

\caption{A correlation of four base-two sequences, which is a base-16 sequence
and an element of $S^{4}(n)$, can typically be visualized as a tetrahedron. These can be interpreted as correlations between two measurements involving separate reference sequences, one associated with Alice (A) and one with Bob (B).}
\label{fig:tetrahedron}       
\end{figure}

The results presented in this paper represent a small fraction of
the modeling potential of this formalism. For example, one can consider higher order correlations between base-2 sequences. In the case of four point correlations between base-2 sequences, the associated geometric elements will typically be tetrahedra (figure \ref{fig:tetrahedron}). Though, unlike three point correlations, there is no guarantee that four randomly selected base-2 sequences will form a valid simplex. This leads to non-trivial behavior of geometric elements beyond two spatial dimensions, which may shed some light on the importance of three spatial dimensions in physics. Each of these geometric elements will have quantum numbers beyond those associated with the lengths of its edges. In the case of four point correlations, there will be ten such quantum numbers. 

Four point correlations between base-2 sequences may also be thought of as two point correlations between base-4 sequences, which we interpret as measurements (figure \ref{fig:tetrahedron}). In other words, we may interpret four point correlations as relationships between two spin measurements, which are necessarily separated in space-time. It is this scenario which corresponds to Stern-Gerlach experiments involving sequences of detectors. Of particular interest are those cases in which two Stern-Gerlach detectors are rotated with respect to one another. A model for this physical scenario will enable us to address the issues of complementarity, as well as the violation of Bell's inequalities \cite{Powers2022}. This, along with the accompanying geometric picture, will also inform the development of a model for space-time. 

A motivating observation of the work presented here is the tension
between the countability of empirical data and the uncountable sets
employed by the theories tasked with modeling that data. What makes
the approach taken here unique is that one need not choose between
these two views of nature. As the length of sequences are taken to
infinity, the number of unique sequences becomes uncountable, leading
to continuous probability distributions. This implies that expectation
values of any observable can then vary continuously, even if that
observable is itself discrete. This feature offers the opportunity
to develop discrete physics models in the finite $n$ regime, while
also studying the continuum limit of those models. This ``continuization'' approach can be contrasted with traditional methods of quantization, which involve the discretization of continuous mathematical structures.

Though the results presented here are promising, there remain many important issues raised in the quantum foundations literature that have not been adequately addressed. These issues include the measurement problem, contextuality, and the role of complex numbers, among many others. Additionally, the precise relationship between various features of the formalism introduced here and QM must still be established. Addressing these issues will, in virtually every case, require a specific model for space-time. Though, the issues concerning quantum foundations will not be the only ones that must be addressed to further justify this research effort. Ultimately, this formalism is only viable if it has the capacity to support both space-time and matter degrees of freedom. For this reason, we must adopt a long term, collaborative approach to model development. The work presented here is intended to establish a general framework upon which such a model can be built. 

The formalism and subsequent model we have introduced are rooted in information theory and have displayed clear predictive power. While these results recast important physics in a new and intriguing light, they are far from the end of the story. There remain important unanswered questions, as well as new questions which we have not yet thought to ask. Given the mathematical simplicity and vivid conceptual picture, we are optimistic that researchers from a broad range of backgrounds will find this effort both enticing and promising.

\section*{Acknowledgments}
We would like to thank Lauren Hay, Omar Elsherif, Rance Solomon, De-Chang Dai, Wei Chen Lin, Will Kinney,
Djordje Minic, Tatsu Takeuchi, Aleksandar Bogojevic, Antun Balaz,
and Aleksandar Belic for comments on previous versions of this work
as well as helpful discussions. We also thank Emily Powers for figure
designs. D.S. is partially supported by the US National Science Foundation, under Grant no. PHY-2014021.

\bibliographystyle{utphys}
\bibliography{references}

\appendix

\section{Derivation of the selection rules for interacting spin systems \label{sec:Derivation-of-AMRA}}

\subsection{Proof of $n=2(j_{10}+g_{10})=2(j_{02}+g_{02})$}

Proving the relation given in equation (\ref{eq: n}) requires us
to add equations (\ref{eq: j}) and (\ref{eq: g}), yielding the following:

\begin{equation}
j+g=\frac{\tilde{C}+\tilde{D}}{2}+\frac{\tilde{A}+\tilde{B}}{2}\label{eq: n step 1-1}
\end{equation}

The length of a particular sequence is given by the total number of
basis elements contained within that sequence. In the case of a base-4
sequence, that is given by $\tilde{A}+\tilde{B}+\tilde{C}+\tilde{D}=n$.
Substituting this result into equation (\ref{eq: n step 1-1}) yields:

\begin{equation}
j+g=\frac{n}{2}\label{eq: n step 2-1}
\end{equation}

An obvious consequence of two base-4 sequences sharing a common base-2
reference sequence, as is the requirement in section \ref{sec:Selection-Rules},
is that both base-4 sequences must be the same length. This fact,
together with equation (\ref{eq: n step 2-1}) yields the result in
equation (\ref{eq: n}):

\begin{equation}
n=2(j_{10}+g_{10})=2(j_{02}+g_{02})\label{eq: n final result-1}
\end{equation}

\subsection{Proof of the selection rules for $m$ and $l$}

The simplest path towards proving equations (\ref{eq: M-1}) and (\ref{eq: L-1})
requires the introduction of the base-2 counts $\tilde{0}_{0}$, $\tilde{1}_{0}$,
$\tilde{0}_{1}$, $\tilde{1}_{1}$, $\tilde{0}_{2}$, and $\tilde{1}_{2}$,
where the subscripts indicate which base-2 sequence each count is
associated with. Using the definition of the base-4 basis elements
$A$, $B$, $C$, and $D$ offered in section \ref{sec:Quantum-Numbers},
the base-4 counts can be expressed in terms of these base-2 counts
like so:

\begin{equation}
\tilde{0}_{0}=\tilde{A}_{10}+\tilde{C}_{10}=\tilde{A}_{02}+\tilde{D}_{02}\label{eq: A1}
\end{equation}

\begin{equation}
\tilde{1}_{0}=\tilde{B}_{10}+\tilde{D}_{10}=\tilde{B}_{02}+\tilde{C}_{02}\label{eq: A2}
\end{equation}

\begin{equation}
\tilde{0}_{1}=\tilde{A}_{10}+\tilde{D}_{10}=\tilde{A}_{12}+\tilde{D}_{12}\label{eq: A3}
\end{equation}

\begin{equation}
\tilde{1}_{1}=\tilde{B}_{10}+\tilde{C}_{10}=\tilde{B}_{12}+\tilde{C}_{12}\label{eq: A4}
\end{equation}

\begin{equation}
\tilde{0}_{2}=\tilde{A}_{02}+\tilde{C}_{02}=\tilde{A}_{12}+\tilde{C}_{12}\label{eq: A5}
\end{equation}

\begin{equation}
\tilde{1}_{2}=\tilde{B}_{02}+\tilde{D}_{02}=\tilde{B}_{12}+\tilde{D}_{12}\label{eq: A6}
\end{equation}

Using equations (\ref{eq:m}), (\ref{eq: A1}) and (\ref{eq: A3}),
the base-4 quantum numbers of interest can be expressed as follows:

\begin{multline}
m_{10}=\frac{\tilde{C}_{10}-\tilde{D}_{10}}{2}=\\
\frac{\tilde{0}_{0}-\tilde{A}_{10}-\tilde{0}_{1}+\tilde{A}_{10}}{2}=\frac{\tilde{0}_{0}-\tilde{0}_{1}}{2}\label{eq: A-inter 1}
\end{multline}

Alternatively, $m_{10}$ can be defined as: 

\begin{multline}
m_{10}=\frac{\tilde{C}_{10}-\tilde{D}_{10}}{2}=\\
\frac{\tilde{1}_{1}-\tilde{B}_{10}-\tilde{1}_{0}+\tilde{B}_{10}}{2}=\frac{\tilde{1}_{1}-\tilde{1}_{0}}{2}\label{eq: A-inter 2}
\end{multline}

By an identical procedure, the quantum number $l_{10}$ can also be
defined in terms of base-2 counts. Generalizing the indices, the following
relations between base-4 quantum numbers and base-2 counts can be
defined:

\begin{equation}
m_{\mu\nu}=\frac{\tilde{0}_{\nu}-\tilde{0}_{\mu}}{2}=\frac{\tilde{1}_{\mu}-\tilde{1}_{\nu}}{2}\label{eq: m gen}
\end{equation}
\begin{equation}
l_{\mu\nu}=\frac{\tilde{0}_{\nu}-\tilde{1}_{\mu}}{2}=\frac{\tilde{0}_{\mu}-\tilde{1}_{\nu}}{2}\label{eq: l gen}
\end{equation}

Using equations (\ref{eq: m gen}) and (\ref{eq: l gen}), with the
appropriate choice of indices, equation (\ref{eq: M-1}) becomes: 

\begin{multline}
m_{12}=m_{10}+m_{02}\rightarrow\\
\frac{\tilde{0}_{2}-\tilde{0}_{1}}{2}=\frac{\tilde{0}_{0}-\tilde{0}_{1}}{2}+\frac{\tilde{0}_{2}-\tilde{0}_{0}}{2}\label{eq: m proof 1}
\end{multline}

\begin{multline}
m_{12}=l_{02}-l_{10}\rightarrow\\
\frac{\tilde{0}_{2}-\tilde{0}_{1}}{2}=\frac{\tilde{0}_{2}-\tilde{1}_{0}}{2}-\frac{\tilde{0}_{1}-\tilde{1}_{0}}{2}\label{eq: m proof 2}
\end{multline}

Equations (\ref{eq: m proof 1}) and (\ref{eq: m proof 2}) both evaluate
to true statements, implying the relations given in equation (\ref{eq: M-1})
are proven. Using equations (\ref{eq: m gen}) and (\ref{eq: l gen}),
with the appropriate choice of indices, equation (\ref{eq: L-1})
becomes: 

\begin{multline}
l_{12}=l_{10}+m_{02}\rightarrow\\
\frac{\tilde{0}_{1}-\tilde{1}_{2}}{2}=\frac{\tilde{0}_{1}-\tilde{1}_{0}}{2}+\frac{\tilde{1}_{0}-\tilde{1}_{2}}{2}\label{eq: l proof 1}
\end{multline}

\begin{multline}
l_{12}=l_{02}-m_{10}\rightarrow\\
\frac{\tilde{0}_{1}-\tilde{1}_{2}}{2}=\frac{\tilde{0}_{0}-\tilde{1}_{2}}{2}-\frac{\tilde{0}_{0}-\tilde{0}_{1}}{2}\label{eq: l proof 2}
\end{multline}

Again, equations (\ref{eq: l proof 1}) and (\ref{eq: l proof 2})
both evaluate to true statements, implying the relations given in
equation (\ref{eq: L-1}) are proven.

\subsection{Proof of the selection rules for $j$ and $g$\label{sec:Derivation-of-Jmax/min}}

As defined in Table \ref{tab:Base-8 measures}, the quantum number
$j_{12}$ can be expressed in terms of base-8 counts like so: 

\begin{equation}
j_{12}=\frac{\widetilde{100}+\widetilde{110}+\widetilde{011}+\widetilde{001}}{2}\label{eq: base-8 J}
\end{equation}

As an explicit example, an element of $S^{3}(n=4)$ is offered, where
brackets around the base-2 basis elements in $s_{j_{1},m_{1},g_{1},l_{1}}^{1}$
and $s_{j_{2},m_{2},g_{2},l_{2}}^{1}$ that contribute to the
quantum numbers $j_{10}$ and $j_{02}$ have been introduced:

\begin{equation}
\left(\begin{array}{ccc}
1 & 1 & [0]\\
1 & 1 & 1\\{}
[1] & 0 & 0\\{}
[0] & 1 & 1
\end{array}\right)\label{eq:Explicit product S^3 example 1}
\end{equation}

In this element of $S^{3}(n=4)$, the bracketed base-2 elements in
$s_{j_{1},m_{1},g_{1},l_{1}}^{1}$ and $s_{j_{2},m_{2},g_{2},l_{2}}^{1}$
do not overlap with one another. This implies that the quantum number
$j_{12}$ between $s_{j_{1},m_{1},g_{1},l_{1}}^{1}$ and $s_{j_{2},m_{2},g_{2},l_{2}}^{1}$
is simply $j_{12}=j_{10}+j_{02}=\frac{(2+1)}{2}=\frac{3}{2}$. On
the other hand, we could have the following situation:

\begin{equation}
\left(\begin{array}{ccc}
[0] & 1 & [0]\\
1 & 1 & 1\\{}
[1] & 0 & 0\\
1 & 1 & 1
\end{array}\right)\label{eq:Explicit product S^3 example 2}
\end{equation}

The difference here is that one of the bracketed base-2 basis elements from
$s_{j_{1},m_{1},g_{1},l_{1}}^{1}$ now overlaps one from $s_{j_{2},m_{2},g_{2},l_{2}}^{1}$.
This implies that the quantum number $j_{12}$ between $s_{j_{1},m_{1},g_{1},l_{1}}^{1}$
and $s_{j_{2},m_{2},g_{2},l_{2}}^{1}$ is now $j_{12}=j_{10}+j_{02}-1=\frac{(2+1)}{2}-1=\frac{1}{2}$.
In other words, given the quantum numbers $j_{10}$ and $j_{02}$,
we can have either $j_{12}=\frac{3}{2}$ or $j_{12}=\frac{1}{2}$.
In general, the allowed range of the quantum number $j_{12}$ is as
follows, which is equation (\ref{eq: J range}):

\begin{equation}
|j_{10}-j_{02}|\leq j_{12}\leq j_{10}+j_{02}\label{eq: beta_max-1}
\end{equation}

In the case that $n<2(j_{10}+j_{02})$, an overlap
is guaranteed. Because $s_{j_{10},m_{10},g_{10},l_{10}}^{2}$ and $s_{j_{02},m_{02},g_{02},l_{02}}^{2}$
share a common reference sequence, the base-4 basis elements that
can overlap in the resulting base-8 sequence are $(A_{10},A_{02})$,
$(B_{10},B_{02})$, $(C_{10},A_{02})$, $(D_{10},B_{02})$, $(A_{10},D_{02})$,
$(B_{10},C_{02})$, $(C_{10},D_{02})$, and $(D_{10},C_{02})$. The
$(C_{10},D_{02})$ and $(D_{10},C_{02})$ cases correspond to the
base-8 basis elements $101$ and $010$ respectively, which are precisely
the overlap scenarios of interest when considering $j_{12}$. Therefore,
the maximum number of overlaps that may occur are limited by the sum
$min\left[\tilde{D}_{10},\tilde{C}_{02}\right]+min\left[\tilde{C}_{10},\tilde{D}_{02}\right]$.
Each overlap leads to a reduction in $j_{12}$ by one, leading to
the following expression:

\begin{multline}
j_{12,min}=j_{10}+j_{02}\\
-min[\tilde{D}_{10},\tilde{C}_{02}]-min[\tilde{C}_{10},\tilde{D}_{02}]\label{eq: j_min}
\end{multline}

$(C_{10},A_{02})$, $(D_{10},B_{02})$, $(A_{10},D_{02})$,
and $(B_{10},C_{02})$ correspond to the base-8 basis elements $100$, $011$,
$001$, and $110$, respectively. This implies that these overlap scenarios all contribute
to $j_{12}$. However, if $n<2(j_{10}+j_{02})$, then it is guaranteed
that either $\tilde{D}_{10}>\tilde{B}_{02}$ or $\tilde{C}_{10}>\tilde{A}_{02}$,
or equivalently $\tilde{B}_{10}<\tilde{C}_{02}$ or $\tilde{A}_{10}<\tilde{D}_{02}$.
This implies that $(C_{10},D_{02})$ and or $(D_{10},C_{02})$ overlap
scenarios must occur. This allows us to define the following expression:

\begin{multline}
j_{12,max}=j_{10}+j_{02}\\
-max\left[0,\tilde{D}_{10}-\tilde{B}_{02}\right]-max\left[0,\tilde{C}_{10}-\tilde{A}_{02}\right]\label{eq: j_max}
\end{multline}

Using the relation between $j$, $g$, and $n$ offered in equation
(\ref{eq: n step 2-1}), the results derived for $j_{12}$ can be used
to derive the corresponding results for $g_{12}$. 

\section{An example of maps\label{sec:Maps}}

Within this formalism, a map connects two sequences of equal basis
and length via element-wise addition modulo two, which is denoted
by the $\oplus$ symbol. That is, given the proper map, any initial
sequence can be mapped to any final sequence like so, where the basis
of these sequences is $2^{d}$:

\begin{equation}
s_{initial}^{d}\oplus s_{map}^{d}=s_{final}^{d}\label{eq: addition example}
\end{equation}

As a more concrete example of the operation shown in equation (\ref{eq: addition example}),
a particular choice of the initial and final sequence is made, where
$s_{initial}^{2}=s_{\frac{1}{2},+\frac{1}{2},\frac{5}{2},+\frac{1}{2}}^{2}$
and $s_{final}^{2}=s_{\frac{1}{2},-\frac{1}{2},\frac{5}{2},-\frac{1}{2}}^{2}$:

\begin{equation}
\left(\begin{array}{c}
A\\
A\\
C\\
B\\
B\\
A
\end{array}\right)\oplus\left(\begin{array}{c}
B\\
A\\
C\\
A\\
A\\
D
\end{array}\right)=\left(\begin{array}{c}
B\\
A\\
A\\
B\\
B\\
D
\end{array}\right)\label{eq: base-4 addition example}
\end{equation}

Expressing these base-4 sequences using the base-2 representation,
we have:

\begin{equation}
\left(\begin{array}{c}
00\\
00\\
10\\
11\\
11\\
00
\end{array}\right)\oplus\left(\begin{array}{c}
11\\
00\\
10\\
00\\
00\\
01
\end{array}\right)=\left(\begin{array}{c}
11\\
00\\
00\\
11\\
11\\
01
\end{array}\right)\label{eq: base-2 addition example}
\end{equation}

The example shown here has the effect of conserving the quantum numbers
$j$ and $g$, but not the quantum numbers $m$ and $l$. Maps which
conserve all quantum numbers are permutations. 

\section{Derivation of summation limits\label{sec:Derivation-of-summation}}

\subsection{Derivation of $k_{min}$ \label{sec:Derivation-of-k0min/max}and
$k_{max}$}

The two overlap scenarios discussed in section \ref{sec:Derivation-of-Jmax/min}
that lead to cases in which $j_{12}<j_{10}+j_{02}$ are $(C_{10},D_{02})$
and $(D_{10},C_{02})$, which correspond to the base-8 basis elements
$101$ and $010$, respectively. In cases where the quantum numbers
$j_{10}$, $j_{02}$, and $j_{12}$ are all known, there still may
be a range of possible values for the counts $\widetilde{010}$ and
$\widetilde{101}$, where the count $\widetilde{010}$ is associated
with the quantum number $k$. It is convenient to introduce the quantum
number $X=k+\widetilde{101}$, where $j_{12}=j_{10}+j_{02}-X$. This
relation implies that for fixed $j_{10}$, $j_{02}$, and $j_{12}$,
the quantum number $X$ is also fixed. Ignoring $X$ for the time
being, we have $\widetilde{101}_{max}=min\left[\tilde{C}_{10},\tilde{D}_{02}\right]$.
For a given $X$, $k_{min}$ must be equivalent to $X-\widetilde{101}_{max}$.
This allows us to define $k_{min}$:

\begin{equation}
k_{min}=max\left[0,X-min\left[\tilde{C}_{10},\tilde{D}_{02}\right]\right]\label{eq: k_0_min}
\end{equation}

Again ignoring $X$, we have $k_{max}=min\left[\tilde{C}_{02},\tilde{D}_{10}\right]$,
which implies the following: 

\begin{equation}
k_{max}=min\left[X,min\left[\tilde{C}_{02},\tilde{D}_{10}\right]\right]\label{eq: k_0_max}
\end{equation}

Thus, given the quantum numbers $j_{10}$, $m_{10}$, $j_{02}$, $m_{02}$,
and $j_{12}$, we can define bounds on the allowed values of $k$. 

\subsection{Derivation of $l_{12,min}$ and $l_{12,max}$}

From Table \ref{tab:Base-8 measures}, the definition of $l_{12}$
in terms of base-8 counts is as follows:

\begin{equation}
l_{12}=\frac{\widetilde{000}+\widetilde{010}-\widetilde{111}-\widetilde{101}}{2}\label{eq: l_12 repeat}
\end{equation}

Given the priors $n$, $j_{10}$, $j_{02}$, $j_{12}$, and a particular
combination of $k^{A}$ and $k^{B}$, along with their definition
in terms of base-8 counts given in Table \ref{tab:Base-8 measures},
the bounds on $l_{12}$ are as follows:

\begin{equation}
l_{12,min}=-\frac{n}{2}+j_{12}+max(k^{A},k^{B})\label{eq: l12min}
\end{equation}

\begin{multline}
l_{12,max}=\frac{n}{2}-j_{12}\\
-max(j_{10}+j_{02}-j_{12}-k^{A},j_{10}+j_{02}-j_{12}-k^{B})\label{eq: l12max}
\end{multline}

\onecolumngrid

\section{The standard closed form Clebsch-Gordan coefficients\label{sec:Expansion-of-the}}

The closed form expression for the Clebsch-Gordan coefficients within
QM takes the following form, where $z$ may take on any value for
which no factorials have negative arguments \cite{Racah1942}:

\begin{multline}
<j_{1}j_{2}JM|j_{1}j_{2}m_{1}m_{2}>=\sqrt{\frac{(2J+1)(j_{1}+j_{2}-J)!(J+j_{1}-j_{2})!(J+j_{2}-j_{1})!}{(j_{1}+j_{2}+J+1)!}}\\
\cdot\sum_{z}(-1)^{z}\frac{\sqrt{(j_{1}+m_{1})!(j_{1}-m_{1})!(j_{2}+m_{2})!(j_{2}-m_{2})!(J+M)!(J-M)!}}{z!(j_{1}+j_{2}-J-z)!(j_{1}-m_{1}-z)!(j_{2}+m_{2}-z)!(J-j_{2}+m_{1}+z)!(J-j_{1}-m_{2}+z)!}\label{eq: Original CGs}
\end{multline}

\end{document}